# A study on Satellite-to-Ground Propagation in Urban Environment

N. Cenni, V. Degli-Esposti, E. M. Vitucci, F. Fuschini, M. Barbiroli

Department of Electrical, Electronic, and Information Engineering "Guglielmo Marconi" (DEI),
CNIT, University of Bologna, 40136 Bologna, Italy

nicolo.cenni@studio.unibo.it, {v.degliesposti, enricomaria.vitucci, franco.fuschini, marina.barbiroli}@unibo.it

*Abstract*— **Non-Terrestrial Networks are going to play an important role in future 6G wireless networks to enhance global connectivity and performances in cooperation with terrestrial networks. In order to properly design and deploy non-terrestrial networks, the satellite-to-ground channel must be properly characterized, with particular focus on the urban environment. This paper uses a Ray-Tracing simulation tool to analyze the primary propagation mechanisms at the ground side and fading characteristics as a function of the satellite position in a reference urban environment. Non-specular reflection due to surface irregularities emerges as a primary propagation mechanism in non-line-of-sight cases. Additionally, the Rician K-factor shows a constant or slightly increasing trend with satellite elevation angle.**

## I. Introduction

Non-Terrestrial Networks (NTN) will be an important element of future 6th generation (6G) wireless networks. Current research is focusing in particular on the development of NTN that can complement terrestrial networks and promote ubiquitous and high-capacity global connectivity. As stated in [1], while previous generation networks have been designed to provide connectivity for a bi-dimensional space, 6G technologies involve a three-dimensional (3D) heterogeneous architecture in which terrestrial infrastructures communicate with non-terrestrial stations such as Unmanned Aerial Vehicles (UAVs), High Altitude Platforms (HAPs) and satellites. The 3D structure is not only able to guarantee adequate broadband coverage to rural and unserved regions and to prevent connectivity outage during unfortunate events such as natural disasters, but also to upgrade the performance of limited terrestrial networks in a cost-effective manner, ensure service availability in high-mobility conditions and reinforce machine to machine communications providing service continuity.

Unfortunately, the number of studies on Satellite-to-Ground (S2G) propagation in urban environment is still limited and standard propagation models such as the 3GPP model [2] appear to be still limited in scope and/or reliability. Looking at the tables of the elevation-dependent Rician K-factor in [2], one may notice some unexplained and counter-intuitive trends of the K-factor, such as:

- Urban and rural scenarios show a completely different trend with respect to dense urban and suburban;
- Suburban S-band show a higher K-factor vs. suburban Ka-band, while all other scenarios show an opposite trend;
- S-band and Ka-band K-factor show almost no difference for the rural scenario, unlike what happens in the other scenarios;
- In the urban case the K-factor decreases very rapidly with elevation, which appears strange.

The latter behaviour might be due to the 3GPP model limitation to Line of Sight (LoS) link conditions, which are far from being the norm in urban environment.

In our work, we focus our attention on the close-to-ground section of the link, to investigate how the signal interacts within the urban environment, while we overlook atmospheric effects and power-budget considerations.

To this aim, we consider an ideal, Manhattan-like urban environment and a full-3D ray tracing tool as a means to better understand the behaviour of S2G propagation in urban environment as a function of satellite elevation and azimuth vs. street orientation. Non-specular scattering, often neglected in many propagation models, is also considered here using the model first presented in [3]. Its impact appears to be crucial, especially when satellite elevation is low and LoS conditions are not satisfied, since both diffraction from the roof and reflection from the nearest walls turn out to be ineffective mechanisms to propagate the signal at street level, as described in [4].

According to our analysis, the K-factor, is found to be constant or to slightly increase with elevation, as intuition would suggest. On the other hand, this simplified topology enabled a comprehensive analysis of propagation mechanisms based on the positions of both satellites and ground stations. Future work will incorporate more realistic urban environment topologies and a more complete propagation characterization.

## II. S2G Simulation Method

In order to investigate propagation and in particular the variation of the K-factor as a function of the elevation angle and the frequency, ray-tracing simulations have been carried out using the advanced 3D ray-tracing simulator presented in [5] which includes diffuse scattering phenomena using the Effective Roughness (ER) model [6].

As a first step, the RT tool has been used for evaluating the contribution of the different propagation mechanisms, namely reflection, diffraction, diffuse scattering and combinations of them, to the total received power, as described in [7], in order to better understand the behaviour of Satellite to Ground (S2G) propagation in urban environment as a function of satellite elevation and azimuth vs. street and set the suitable RT parameters.

To this aim an ideal, Manhattan-like urban environment has been considered with 40x30 m city blocks and 20 m wide streets (Figure 1). The height of the buildings of 20 m was chosen according to [8]. The frequency is set at 3 GHz, therefore in the S-band and isotropic antennas are assumed both at the Tx and Rx sides, in order to make the outcomes more general and not related to a specific satellite configuration. Considering the satellite side, even if the satellite beam has a certain directivity, the footprint of the satellite on ground affects in the same way all the receiver locations, as simulations have been performed over an area of 300x100 m, therefore, the satellite illumination is uniform over the area. This choice is made also not to account for orientations of the receiving antennas in case of directive antennas.

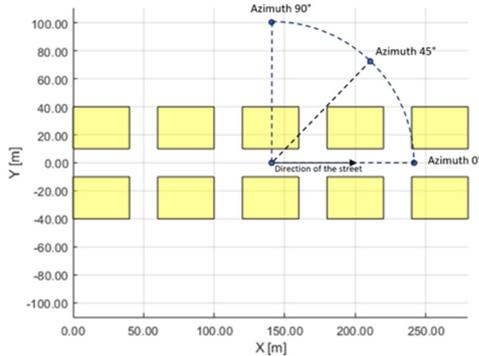

Figure 1. Manhattan-like scenario for RT simulations

Table I summarise the RT parameters used in RT simulations.

*Table I – Simulation parameters.*

| Ray Tracing simulation parameters | |
|---|---|
| Tx elevation angles | 10°,20°,30°,40°,50°,60°, 70°,80°,90° |
| Tx azimuth w.r.t. street orientation | 0°, 45°,90° |
| Tx and Rx antennas | isotropic |
| Frequency | 3 GHz (S-band) |
| Scattering parameter | S = 0.4 |
| Tile area for scattering | 25 m² |
| Building relative dielectric | $\varepsilon_r = 5$ |
| Building conductivity | $\sigma = 0.01$ |
| Scattering model | directive |

In addition to the analysis of the dominant propagation mechanisms, we carried out the analysis of the small-scale fading assuming Rician fading statistics, as in the in the 3GPP model [2]. It is known that the time-space varying envelope of the received signal in presence of a dominant path, such as the Line of Sight (LoS) path, can be described by a Rician distribution. The Rician K-factor can be defined as the ratio of the dominant component power over the (local-mean) scattered power.

In this study, the K-factor for different elevation and azimuth angles of the transmitter and for different receiver positions has been extracted. Specifically, for each position of the receivers on the map a 15x15 grid of the order of few wavelengths is considered, in order to sample fast fading (Figure 2). Then a maximum likelihood estimator is applied to estimate the parameters of the Rice distribution and therefore the K-factor.

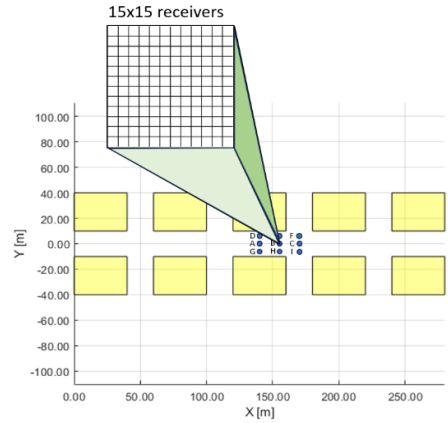

Figure 2. Manhattan-like scenario for RT simulations

### III. RESULTS AND DISCUSSION

RT simulations have been carried out considering receivers located in the different positions of the map, and several elevation angles (from 10° to 90°) and azimuth angles (from 0° to 90°) of the transmitter with respect to the street orientation. The types of propagation mechanisms that have been considered in the simulation are:

- LoS (abbreviation L in the legends);
- Reflections, up to a maximum of 4 (abbreviation R);
- Diffractions, up to a maximum of 3 (abbreviation D);
- Non-specular diffuse scattering (abbreviation S);
- Reflections combined with diffractions, up to a maximum of 4 interactions in total (abbreviation RD);
- Reflections combined with scattering, up to a maximum of 2 (abbreviation RS);
- Diffractions combined with scattering, up to a maximum of 1 (abbreviation DS).

In the RT simulations, diffuse scattering rays are generated by dividing each wall into "tiles" with 5x5 m2 area: non specular-rays are traced from the center of each tile, and the field amplitude is computed according to the ER model described in [6]. Specularly reflected and diffracted rays are traced using the image method, and their field is computed according to the Fresnel's and

Uniform Theory of Diffraction (UTD) coefficients, respectively.

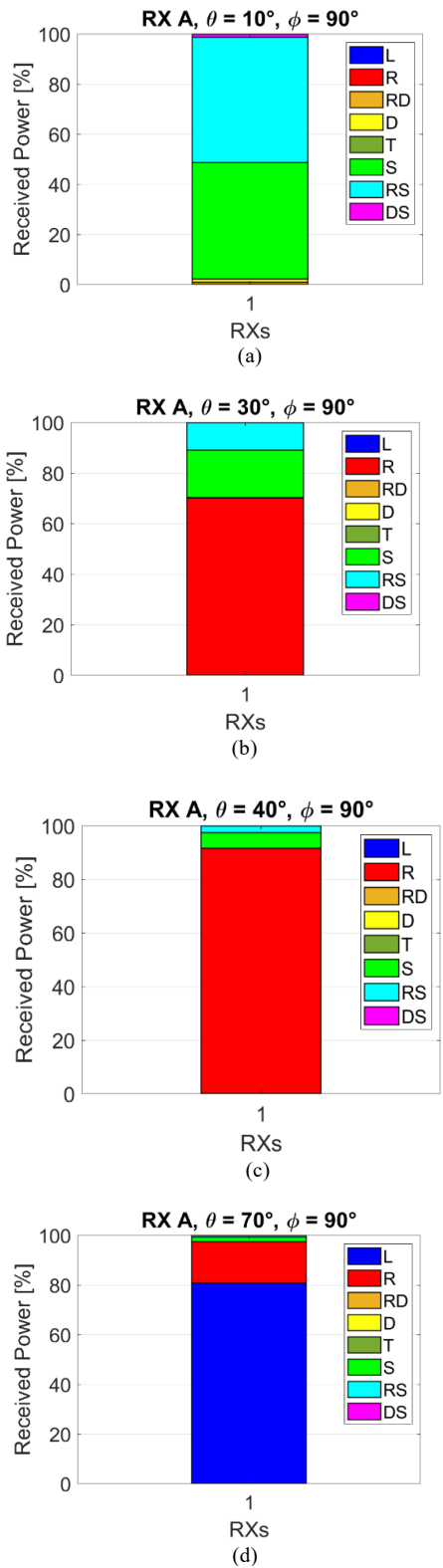

Figure 3. Bar graphs of the percentage of the total received power associated to each propagation mechanism for elevation angles of 10°, 30° 40° and 70°, and azimuth angle 90°, for a single receiver.

In Figure 3(a), 3(b), 3(c) and 3(d) the results for few significant elevation angles of the transmitter (θ=10°, 30°, 40°, 70°) and a fixed azimuth angle (ϕ=90°) are presented, when the receiver is positioned in the center of the map (in the middle of the street and far from the street intersection). The plots show in different colours the percentage contribution of each propagation mechanism, and their combinations, on the total received power.

For small elevation angles, single scattering and scattering combined with reflection are the predominant mechanisms, as stated before. Then, as the elevation approaches 30° the reflection contribution is increasingly more observable as reflections on the street canyon with 2 or 3 bounces are able to reach the receiver. Moreover, by further increasing the elevation angle up to 40°, specular reflection becomes the undisputed dominant mechanism as it is possible to reach the receiver through a single reflection bounce. Between 60° and 70° of elevation, the LoS path appears and clearly becomes the new dominant mechanism, followed by reflection from the ground, whereas reflections from building walls cease to exist.

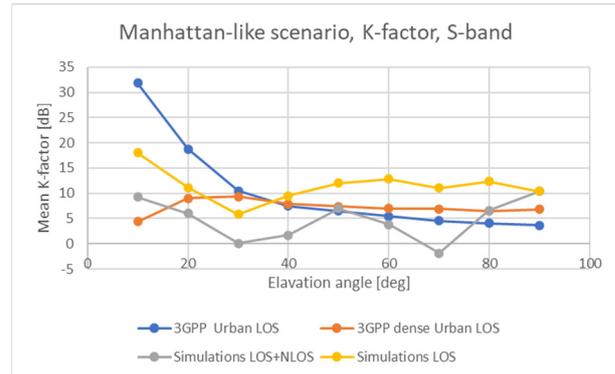

Figure 4. Comparison between the K-factor results, averaged over spatial position and azimuth angles and 3GPP Urban and Dense Urban K-factor values.

In Figure 4 the K-factor in dB, extracted with the RT tool and averaged over several spatial positions and azimuth angles, is shown as a function of the Tx elevation angle. In the figure, the results for LoS-only configurations (yellow line), and for locations including both LoS and NLoS configurations (grey line) are separately shown and compared with the values reported in [2] for the "urban" and "dense urban" environments, in the case of LoS-only configurations (blue and orange line respectively).

From Figure 4 the K-factor generally appears to slightly increase with elevation, mainly if we consider both LoS and NLos locations, and this trend seems more physically sound than the decreasing trend of the 3GPP model in urban environment (blue line), as for lower elevations the multipath richness is generally larger. One of the reasons for this discrepancy may be that simulations ran for deriving parameters in [2] neglect fundamental propagation mechanisms such as diffuse scattering, which is able to accounts for interaction of scatterers near the receiver, including urban furniture and cars. Moreover, the whole set of receiver points (LoS and NLoS) has been considered not only to increase the statistical power but also to observe the transition of the K-factor behavior as the receivers shifts from NLOS to LOS conditions.

This preliminary investigation aimed at assessing the methodology and, based on the outcomes of the propagation mechanism analysis, the RT parameters to be used in larger simulations in real environment. Further studies dealing with the analysis of S2G propagation as a function of a larger number of parameters such as scattering-tile dimensions, and in more realistic scenarios (dense urban, urban and suburban) have to be carried out for an overall assessment of the K-factor.

## IV. Conclusions

In this paper an analysis of propagation in an ideal, satellite to ground urban environment is carried out using a 3D-RT simulation tool. In particular, the Rician K-factor as a function of the satellite elevation angle is evaluated. The RT tool also accounts for non-specular scattering from building surfaces, which is shown to be a fundamental propagation mechanism especially when visibility between Tx and Rx is not present. Simulations show that diffuse scattering is the main propagation mechanism for non-line-of-sight locations, then for increasing values of the elevation angles specular reflection become a dominant mechanism, until when direct visibility is reached, and the line-of-sight contribution clearly becomes dominant.

Additionally, the analysis of the Rician K-factor shows a slightly increasing trend with increasing elevation angles. The high K-factor values we found, however, should be confirmed with other simulations in more realistic environments and with larger set of simulation data.

## V. Acknowledgement

This work was supported in part by the European Union under the Italian National Recovery and Resilience Plan (NRRP) of NextGenerationEU, a partnership on "Telecommunications of the Future" (PE00000014 - program "RESTART"), and in part by the EU COST Action INTERACT (Intelligence-Enabling Radio Communications for Seamless Inclusive Interactions), Grant CA20120.

## References

[1] M. Giordani and M. Zorzi, "Non-terrestrial networks in the 6G era: Challenges and opportunities," *IEEE Network*, vol. 35, no. 2, pp. 244–251, 2021.

[2] 3GPP, "Study on new radio (NR) to support non-terrestrial networks: TR 38.811 v15. 4.0," 2020.

[3] V. Degli-Esposti, F. Fuschini, E. M. Vitucci, G. Falciasecca, "Measurement and modelling of scattering from buildings", *IEEE Trans. Antennas Propag.*, Vol. 55 No 1, pp. 143-153, Jan. 2007.

[4] V. Degli-Esposti, F. Fuschini, and D. Guiducci, "A study on roof-to-street propagation," IEEE 2003 International Conference on Electromagnetics in Advanced Applications (ICEAA '03), Turin, Italy, 8-12 September 2003.

[5] F. Fuschini, E. M. Vitucci, M. Barbiroli, G. Falciasecca, V. Degli-Esposti, "Ray tracing propagation modeling for future small-cell and indoor applications: a review of current techniques," *Radio Science*, Vol. 50, No. 6, pp. 469–485, June 2015.

[6] E. M. Vitucci, N. Cenni, F. Fuschini, V. Degli Esposti, "A Reciprocal Heuristic Model for Diffuse Scattering from Walls and Surfaces," *IEEE Trans. Antennas Propag.*, vol. 71, no. 7, pp. 6072-6083, Jul. 2023.

[7] L. Possenti, M. Barbiroli, E. M. Vitucci, F. Fuschini, M. Fosci and V. Degli-Esposti, "A Study on mm-Wave Propagation in and Around Buildings," *IEEE Open Journal Antennas Propag.*, vol. 4, pp. 736-747, 2023, doi: 10.1109/OJAP.2023.3297201.

[8] Report ITU-R M.2135, "Guidelines for evaluation of radio interface technologies for IMT-Advanced," 2008.